\documentstyle[psfig]{aipproc}

\begin{document}

%
\def\sol {$_\odot~$}
\def\etal {{\it et~al.~}}
\def\eol     {\hfil\break}
\def\endpage {\vfill\supereject}
\newbox\grsign \setbox\grsign=\hbox{$>$} 
\newdimen\grdimen \grdimen=\ht\grsign
\newbox\laxbox \newbox\gaxbox
\setbox\gaxbox=\hbox{\raise.5ex\hbox{$>$}\llap
     {\lower.5ex\hbox{$\sim$}}}\ht1=\grdimen\dp1=0pt
\setbox\laxbox=\hbox{\raise.5ex\hbox{$<$}\llap
     {\lower.5ex\hbox{$\sim$}}}\ht2=\grdimen\dp2=0pt
\def\gax{\mathrel{\copy\gaxbox}}
\def\lax{\mathrel{\copy\laxbox}}

\title{BATSE Evidence for \\ GRB Spectral Features}

\author{
M. S. Briggs$^*$,
D. L. Band$^\dagger$,
R. D. Preece$^*$,
G. N. Pendleton$^*$,  
W. S. Paciesas$^*$ \&
J. L. Matteson$^\dagger$}

\address{
$^*$Department of Physics, University of Alabama in Huntsville, \\
$^\dagger$Center for Astrophysics and Space Science, University
of California, San Diego}

\maketitle

\begin{abstract}
We have developed an automatic search procedure to identify low-energy
spectral features in GRBs.    We have searched
133,000 spectra from 117 bright bursts and have identified 12 candidate
features with significances ranging from
our threshold of $P=$ 5E$-$5 to $P=$ 1E$-$7.
Several of the candidates have been examined in detail, including some
with data from more than one BATSE spectroscopy detector.
The evidence for spectral features appears good; however, the features
have not conclusively been shown to be narrow lines.
\end{abstract}

\section*{Line Search and Results}

Narrow, low-energy ($< 100$ keV) spectral lines in GRBs have been
reported using the data of several instruments (e.g., see review \cite{my_rev}).
Based on these reports,                                  
the BATSE team expected lines to be easy to find and looked
for them manually \cite{palmer}.
The reality was different: lines were not obvious in burst spectra.
In order to be sure that the manual search had not missed any lines,
we implemented a comprehensive, automatic computer search \cite{ls}.
Bursts with at least one spectrum with a normed signal-to-noise ratio
(SNR) \cite{det_prob} near 40~keV
above 5.0 are searched.
For each burst, we form spectra from each individual spectral record,
every pair, triple, etc.
The spectra so formed overlap in many cases.
Once a burst is selected for the search,
spectra are searched regardless of the presence of burst flux---low SNR
spectra serve as controls.
Each spectrum is fit with a continuum model and then a series of fits
are made adding narrow lines
at a closely spaced grid of line centroids extending to 100 keV.
Line candidates are identified by a $\chi^2$ change 
$\Delta \chi^2$ of more than 20,
corresponding to a chance probability in a single spectrum of
$P = 5\times 10^{-5}$.
The probability is calculated for two line parameters, intensity
and centroid, since the intrinsic width is assumed narrow compared to
the detector resolution.

So far, 133,000 spectra formed from 12,000 spectral records
from 117 bursts have been searched.
Most of these spectra  have SNR too low to support the detection
of a line.    Only 16,000 have normed SNR $>$ 5, which our simulations show
is needed to have a reasonable sensitivity to lines similar to those
found in the {\it Ginga} data \cite{BATSE_GL}.

The search identified several cases with $\Delta \chi^2$ values
exceeding our significance threshold, but which we rejected
either because they occur in
background intervals or because they become insignificant when
a better fit requiring  a non-negative flux is made.
These cases were located in spectra of low SNR, and 
because of the large number of low-SNR spectra, they
are consistent with statistical fluctuations.
Several possible candidates with angles between the burst direction and detector
normal exceeding $70^\circ$ are inconsistent with the data of
other detectors.    We believe these cases to be due to
inadequacies in the detector model and we have set them aside.

After these exclusions,  our candidate list has 
12 members with $\Delta \chi^2$
values ranging from the threshold of 20 to 31.8 ($P = 1 \times 10^{-7}$).
The normed SNRs of the spectra in which these candidates
are most significant range from 2.1 to 18.4, with 10
normed SNR values exceeding 5.
These SNR values are reasonable for real features.

The number of independent trials is a nebulous concept.
While there are roughly 16,000 spectra sufficiently bright 
for there to be
a high probability of detecting a real line,
many of these spectra overlap, e.g.,
starting a few records earlier or later, or differing
in length by a few records. 
Consequently the number of bright, independent spectra is one or more
orders of magnitude below 16,000.
The number of independent energy resolution elements averages about
5 per spectrum.
We estimate that the ensemble chance probability of the most-significant
candidate is below $1 \times 10^{-3}$, and probably much lower than
this value.    

We therefore
believe that few, if any, of these candidates are statistical fluctuations.
Of the 12 candidates, in all but one
the best line fit is an emission feature
with a centroid near 40~keV.
The twelfth candidate is an absorption feature at 60~keV.
Typically the lowest energy of the data is about 20 keV,
so we are unable to find lines much below 40~keV.

\section*{Example Candidate}

We have previously presented results \cite{ls} for a candidate in
GRB~940703 (trigger 3057) which
was usefully observed by only one BATSE spectroscopy detector (SD).
Observations are useful only from SDs in high-gain mode which point
to within $70^\circ$ of the burst.
Since that time we have searched more bursts and begun more
detailed analyses of the nine candidates that were usefully
observed by more than one SD.

These multiple-detector cases allow stringent tests of the reality
of the candidates.
Ideally, the data from another detector will confirm
the feature found by the automatic search.
While some features may not be confirmed by another detector
due to differing SNRs between the detectors or from plausible
statistical fluctuations,
the data of all detectors must be consistent.

\begin{figure}[tb!]
\mbox{
\psfig{file=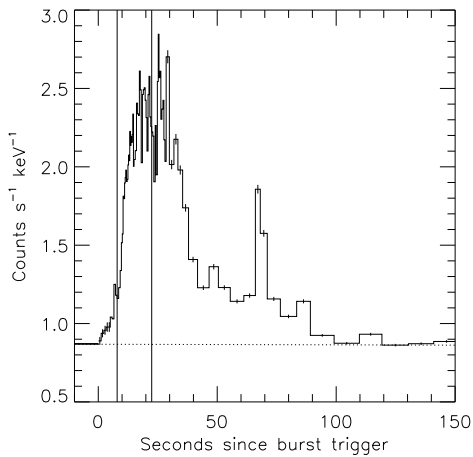,height=2.5in,%
bbllx=40bp,bburx=530bp,bblly=180bp,bbury=666bp,clip=}
\hspace{2mm}
\begin{minipage}[b]{67mm}
\caption{The~time history of GRB~941017 as observed with SD~0.
The automatic search found the greatest $\Delta \chi^2$ value, 20.8 
($P = $ $3 \times 10^{-5}$), for
adding a narrow line for the interval 7.936 to 22.464~s, which
is designated with the pair of vertical lines.
The dotted line indicates the background model.
} 
\protect \vspace{13mm}
\end{minipage}
}
\end{figure}

GRB 941017 (trigger 3245) was usefully observed by SDs 0 and~5.
The automatic search identified a candidate in the data of SD~0 
during the rising portion of the burst (Fig. 1).
A spectrum is not available for SD~5 for the time interval with
the best $\Delta \chi^2$ value for SD~0.
We therefore continued our analysis using spectra from a similar time interval
(9.728 to 23.936~s) which is
available for both detectors.
The results of fits to this interval are listed in Table~1.
Because of the non-optimum time interval, the significance of the
feature in the data of SD~0 is reduced.
The line was discovered in the data of SD~0 so we may regard the centroid
as a priori determined when we analyze the data of SD~5.
A one parameter
line fit yields a significance for the feature in SD~5 of
$4 \times 10^{-3}$.
The joint fit (Fig.~2) yields a significance of $2 \times 10^{-6}$,
better than the value obtained with SD~0 alone
despite the forced
change to the common time interval.
SD~5 provides evidence for the feature independent 
of the detector in which the feature was found by the
automatic search.

\begin{table}[tb!]
\begin{center}
\caption{Line fits to the common time interval, GRB~941017.}
\begin{tabular}{cccccc}
Detectors  &  $\Delta \chi^2$  &   $P$    &   \# of    &   line intensity  &
centroid   \\
Fit  &   &   & line param.  &  ($\gamma$ s$^{-1}$ cm$^{-2}$)   \\
\tableline
0  &  19.1  &   7 $\times 10^{-5}$   &   2   & 0.31$\pm$0.07  &   43.8$\pm$1.2   \\
5  &   9.3  &   1 $\times 10^{-2}$  &   2   & 0.21$\pm$0.07  &   42.2$\pm$1.7  \\
5  &   8.5  &   4 $\times 10^{-3}$  &   1   & 0.19$\pm$0.07  &   43.8  \\
 0 \& 5 &  26.5 &  2 $\times 10^{-6}$  &  2   &  0.25$\pm$0.05  &  43.1$\pm$1.0  \\
\end{tabular}
\end{center}
\end{table}

\begin{figure}[tb!]
\mbox{
\psfig{file=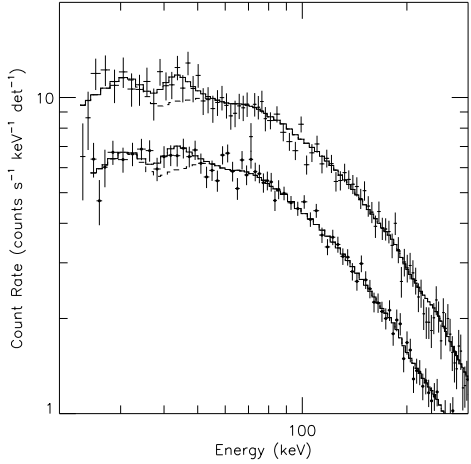,height=2.5in,%
bbllx=40bp,bburx=535bp,bblly=180bp,bbury=666bp,clip=}
\hspace{3mm}
\begin{minipage}[b]{66mm}
\caption{Simultaneous fit to the data for GRB~941017 from SD~0 
(crosses) and SD~5 (crosses with dots---shifted
downwards by $\times 5$ for clarity).
The photon model
has been folded through the detector responses to
produce the count rate models (histograms).
The dashed line shows the continuum portion of the model.
The width of the feature in count space is due to the resolution of
the detector.
The `hump' at 30~keV is due to iodine
in the detector \protect \cite{my_rev}. }
\protect \vspace{7 mm}
\end{minipage}
}
\end{figure}

We must also check that the data of the two detectors are consistent.
For example, are the line strengths or significances obtained from each
detector compatible with the values obtained with the other detector?
We expect some differences due to statistical fluctuations.
Therefore we have investigated this question with simulations based
upon the parameters of the joint fit.
Table~2 shows the data of  the two detectors to be consistent.

\begin{table}[bt!]
\begin{center}
\caption{Simulation Results: Tail-probabilities of the fits to each
detector assuming the parameters of the joint fit.}
\begin{tabular}{lcc}
Fraction with a greater line intensity:  &   SD 0:   &  0.28  \\
                                         &   SD 5:   &  0.85  \\
Fraction with a greater $\Delta \chi^2$: &   SD 0:   &  0.24 \\
                                         &   SD 5:   &  0.88  \\
\end{tabular}
\end{center}
\end{table}

To demonstrate the existence of a line, we must show that the data require
a line independent of which reasonable continuum model is assumed
\cite{fen88,my_rev}.
The above fits were made using the standard `GRB' continuum function of
Band.
However, if we use an alternative continuum model
with a low-energy break
{\it in addition} to the standard high-energy break,
the significance of the line in SD~0 is reduced to
$\Delta \chi^2 = 6.3$   ($P$= 4\%).
While this model still involves a low-energy spectral feature (Fig.~3),
it means
that for this burst we cannot prove that the feature is a line.

\begin{figure}[tb!]
\mbox{
\psfig{file=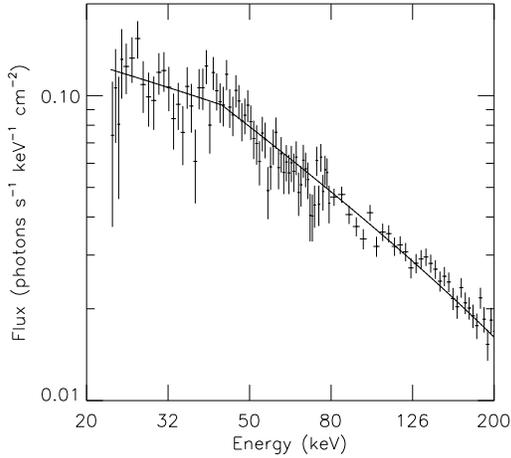,height=2.5in,%
bbllx=40bp,bburx=552bp,bblly=180bp,bbury=660bp,clip=}
\hspace{3mm}
\begin{minipage}[b]{63mm}
\caption{Fit to the data of SD~0, GRB~941017 using an alternative continuum
model.
The continuum model used looks very similar to 
the Band `GRB' form---in $\nu F_{\nu}$ space, the energy emission
peaks at 600 keV. 
To this model we have added an additional break at 43 keV, for a total
of 6 parameters (vs. the 4 parameters of the GRB function).
The fit has $\chi^2 = 201$ for 189 degrees-of-freedom.
The depiction of the
data points in photon flux units is model-dependent.}
\protect \vspace{4mm}
\end{minipage}
}
\end{figure}


\section*{Conclusions}

Since our last report \cite{ls}, we have completed the automatic search for
bright bursts through May 1996.
We have identified 12 line candidates and have begun
detailed analysis.
Because the 12 candidates appear as expected in the small fraction of
spectra with high SNR,  their probability of appearing in the ensemble
by chance is low.

For all except one of the 12 candidates, the best line fit is an
emission line with a centroid near 40~keV.
Our simulations show that we have useful sensitivity to {\it Ginga}-like
absorption lines at 40~keV \cite{BATSE_GL,BAND_conf},
so the lack of detections of absorption lines
gives a constraint on their frequency.
However, with the small number of bright bursts seen by BATSE and {\it Ginga},
there is no significant discrepancy between the two instruments
\cite{det_prob}.

Observations made with multiple detectors have the
strong advantage of redundantly verifying the existence of the features.
We have now carefully investigated several candidate features, including
some observed with multiple detectors.  So far
the evidence appears good for the existence of spectral features,
however we do not want to make a blanket statement that BATSE has observed
spectral features until we have carefully examined all 12 candidates.

Note that we are stating that the
evidence appears good for the existence of {\it spectral features.}
In the cases examined so far, we have not been able to demonstrate
that the spectral features must be narrow lines---an alternative 
explanation is possible: a low-energy break in addition to the normal
break in the few 100 keV to 1 MeV region.

\end{document}